\documentclass[12pt]{article}
\newcommand{\bq}{\begin{equation}}
\newcommand{\eq}{\end{equation}}
\newcommand{\ba}{\begin{eqnarray}}
\newcommand{\ea}{\end{eqnarray}}

\newcommand{\nl }{ \nonumber  }

\newcommand{\p}{\partial}
\newcommand{\h}{\hspace{.5cm}}
\newcommand{\s}{\sigma}

\begin{document}
\begin{center}
{\bf ON THE CLASSICAL STRING SOLUTIONS AND STRING/FIELD THEORY
DUALITY \vspace*{0.5cm}\\ D. Aleksandrova, P. Bozhilov}
\\ {\it Department of Theoretical and Applied Physics, \\
Shoumen University, 9712 Shoumen, Bulgaria\\
E-mail:} p.bozhilov@shu-bg.net
\end{center}
\vspace*{0.5cm}

We classify almost all classical string configurations, considered
in the framework of the semi-classical limit of the string/gauge
theory duality. Then, we describe a procedure for obtaining the
conserved quantities and the exact classical string solutions in
general string theory backgrounds, when the string embedding
coordinates depend non-linearly on the worldsheet time parameter.


\vspace*{.5cm}

\section{Introduction}
After the appearance of the article \cite{4} on the semi-classical
limit of the gauge/string correspondence, a lot of papers have
been devoted to the investigation of the connection between the
classical string solutions, their semi-classical quantization and
the string/field theory duality \cite{5}, \cite{7} - \cite{9},
\cite{11} - \cite{17}, \cite{19} - \cite{29}. Different string
configurations have been considered: rotating, pulsating and
orbiting strings. Most of the authors investigated the closed
string case. In \cite{22,26} the open string case was considered,
where nontrivial boundary conditions had to be also satisfied.

The most explored background was $AdS_5\times S^5$. However, the
string dynamics has been investigated in many other string theory
backgrounds, known to have field theory dual descriptions in
different dimensions, with different number of (or without)
supersymmetries, conformal or non-conformal. The influence of the
$B$-field on the obtained string solutions has been also
considered. Besides, solutions for higher dimensional extended
objects (M2-, D3- and M5-branes) have been obtained
\cite{6,10,14,27,18}.

For establishing the correspondence between the semi-classically
quantized string solutions and the appropriate objects in the dual
field theory, it is essential for one to know the explicit
expressions for the conserved charges like energy, spin, etc., on
the string theory side. Their existence is connected with the
symmetries of the corresponding supergravity backgrounds,
generated by the Killing vectors for these backgrounds. The
analysis of the connection between the ansatzes, used to obtain
exact string solutions, and the background symmetries shows that
the latter have been essentially explored in the process of
solving the string equations of motion and constraints.

The comparison of the ansatzes used in \cite{4,5}, \cite{7} -
\cite{9}, \cite{11} - \cite{17}, \cite{19} - \cite{29}, with the
corresponding symmetries of the target space-times, led us to the
following classification scheme: \ba\label{tLA} X^\mu (\tau,
\sigma)= \Lambda^\mu_0\tau + \Lambda^\mu_1\sigma,\h X^a (\tau,
\sigma)= Y^a (\tau);\\ \label{tGA} X^\mu (\tau, \sigma)=
\Lambda^\mu_0\tau + \Lambda^\mu_1\sigma + Y^\mu (\tau),\h X^a
(\tau, \sigma)= Y^a (\tau);\\ \label{sLA} X^\mu (\tau, \sigma)=
\Lambda^\mu_0\tau + \Lambda^\mu_1\sigma,\h X^a (\tau, \sigma)= Z^a
(\sigma);\\ \label{sGA} X^\mu (\tau, \sigma)= \Lambda^\mu_0\tau +
\Lambda^\mu_1\sigma + Z^\mu (\sigma),\h X^a (\tau, \sigma)= Z^a
(\sigma);\\ \nl \Lambda^\mu_m = const,\h (m=0,1).\ea In the above
equalities, the string embedding coordinates $X^M(\tau,\sigma)$,
$(M=0,1,\ldots, D-1)$, are divided into coordinates
$X^\mu(\tau,\sigma)$ and $X^a(\tau,\sigma)$, such that \ba\nl
dim\{\mu\} + dim\{a\}= dim\{M\}.\ea $X^\mu(\tau,\sigma)$
correspond to the space-time coordinates $x^\mu$, on which the
background fields do not depend. In other words, there exist
$dim\{\mu\}$ commuting Killing vectors $\p/\p x^\mu$. In
(\ref{tLA}) - (\ref{sGA}), we have separated the cases $Y^\mu =0$
and $Y^\mu\ne 0$, $Z^\mu =0$ and $Z^\mu\ne 0$, because the types
of the solutions in these cases are essentially different, as we
will see later on.

All the ansatzes used in \cite{4,5}, \cite{7} - \cite{9},
\cite{11} - \cite{17}, \cite{19} - \cite{29}, are particular cases
of (\ref{tLA}) - (\ref{sGA}), except two \cite{21}. In
\cite{11,14,19,28} there are of type (\ref{tLA}), in \cite{4} and
\cite{12} - of type (\ref{tGA}). In \cite{4,5}, \cite{7} -
\cite{9}, \cite{12}, \cite{13}, \cite{27}, \cite{15}, \cite{19} -
\cite{25}, \cite{29}, ansatzes of the type (\ref{sLA}) are used.
Solutions, based on the ansatzes of the type (\ref{sGA}), are
obtained in \cite{16,17,19,26}.

The aim of this article is to describe a procedure for obtaining
the conserved quantities and the exact classical string solutions
in general string theory backgrounds, based on the ansatzes
(\ref{tLA}) and (\ref{tGA}). Besides, we will use more general
worldsheet gauge than the conformal one, in order to be able to
discuss the tensionless limit $T\to 0$, corresponding to small
t'Hooft coupling $\lambda\to 0$ on the field theory side. The zero
tension limit is also interesting in connection with the ongoing
discussion on massless higher spin field theories.

The paper is organized as follows. In Sec.2 and Sec.3, we describe
the string dynamics and obtain the corresponding exact solutions,
based on the ansatzes (\ref{tLA}) and (\ref{tGA}) respectively.
Sec.4 is devoted to some applications of the obtained general
results.

\section{Exact string solutions}
The Polyakov type action for the bosonic string in a
$D$-dimensional curved space-time with metric tensor $g_{MN}(x)$,
interacting with a background 2-form gauge field $b_{MN}(x)$ via
Wess-Zumino term, can be written as \ba\label{pa}
&&S^{P}=-\frac{1}{2}\int
d^{2}\xi\Bigl\{T\sqrt{-\gamma}\left[\gamma^{mn} \p_m X^M\p_n X^N
g_{MN}\right]-Q\varepsilon^{mn} \p_{m}X^{M}\p_{n}X^{N}
b_{MN}\Bigr\},\\ \nl && \p_m=\p/\p\xi^m,\h
\xi^m=(\xi^0,\xi^1)=(\tau,\s),\h m,n = 0,1;\h M,N =
0,1,\ldots,D-1,\ea where $\gamma$ is the determinant of the
auxiliary worldsheet metric $\gamma_{mn}$, and $\gamma^{mn}$ is
its inverse. The position of the string in the background
space-time is given by $x^M=X^M(\xi^m)$, and $T=1/2\pi\alpha'$,
$Q$ are the string tension and charge, respectively. If we
consider the action (\ref{pa}) as a bosonic part of a
supersymmetric one, we have to set $Q=\pm T$. In what follows, $Q
=T$.

The action (\ref{pa}) is known to be classically equivalent to the
Nambu-Goto type action: \ba\nl S^{NG}=- T\int d^{2}\xi\left[
\sqrt{-G}-\frac{1}{2}\varepsilon^{mn} \p_{m}X^{M}\p_{n}X^{N}
b_{MN}(X)\right],\ea where $G\equiv\det(G_{mn})$ and \ba\nl
G_{mn}=\p_m X^M\p_n X^N g_{MN}(X)\ea is the metric induced on the
string worldsheet. We will work with the Polyakov type action.

The equations of motion for $X^M$ following from (\ref{pa}) are:
\ba \label{em}
&&-g_{LK}\left[\p_m\left(\sqrt{-\gamma}\gamma^{mn}\p_nX^K\right) +
\sqrt{-\gamma}\gamma^{mn}\Gamma^K_{MN}\p_m X^M\p_n X^N\right]\\
\nl &&=\frac{1}{2}H_{LMN}\epsilon^{mn}\p_m X^M\p_n X^N,\ea where
\ba\nl
&&\Gamma_{L,MN}=g_{LK}\Gamma^K_{MN}=\frac{1}{2}\left(\p_Mg_{NL}
+\p_Ng_{ML}-\p_Lg_{MN}\right),\\ \nl &&H_{LMN}= \p_L b_{MN}+ \p_M
b_{NL} + \p_N b_{LM}.\ea The constraints are obtained by varying
the action (\ref{pa}) with respect to $\gamma^{mn}$:
\ba\label{con} \delta_{\gamma^{mn}}S^P=0\Rightarrow \left(\p_m
X^M\p_n X^N-\frac{1}{2}\gamma_{mn}\gamma^{kl} \p_k X^M\p_l
X^N\right)g_{MN}=0.\ea

Now, we would like to solve (\ref{em}) and (\ref{con}). Let us
first consider the constraints (\ref{con}). In order to work with
$\gamma^{mn}$ only, we rewrite them as \ba\label{oc}
\left(\gamma^{kl}\gamma^{mn}-2\gamma^{km}\gamma^{ln}\right)G_{mn}=0.\ea
We have three constraints in (\ref{oc}), but only two of them are
independent. To extract the independent ones, we rewrite the three
constraints as follows: \ba\label{00}
&&\left(\gamma^{00}\gamma^{mn}-2\gamma^{0m}\gamma^{0n}\right)G_{mn}=0,\\
\label{01}
&&\left(\gamma^{01}\gamma^{mn}-2\gamma^{0m}\gamma^{1n}\right)G_{mn}=0,\\
\label{11}
&&\left(\gamma^{11}\gamma^{mn}-2\gamma^{1m}\gamma^{1n}\right)G_{mn}=0.\ea
Inserting $G_{00}$ from (\ref{00}) into (\ref{01}) and (\ref{11}),
one obtains that both of them are satisfied, when the equality
\ba\label{1} \gamma^{00}G_{01}+\gamma^{01}G_{11}=0 \ea is
fulfilled. To simplify the constraint (\ref{00}), we put (\ref{1})
in it, which results in \ba\label{0}
\gamma^{00}G_{00}-\gamma^{11}G_{11}=0.\ea So, our {\it
independent} constraints, with which we will work from now on, are
given by (\ref{1}) and (\ref{0}).

Now let us turn to the equations of motion (\ref{em}). We will
work in the gauge $\gamma^{mn}=constants$, in which they simplify
to \ba\label{sem} -\sqrt{-\gamma}\gamma^{mn}
g_{LK}\left[\p_m\p_nX^K +\Gamma^K_{MN}\p_m X^M\p_n
X^N\right]=\frac{1}{2}H_{LMN}\epsilon^{mn}\p_m X^M \p_n X^N .\ea
In particular, $\gamma^{mn}=\eta^{mn}=diag(-1,1)$ correspond to
the the usually used {\it conformal gauge}.

\subsection{Solving the equations of motion and constraints}
In the frequently used {\it static gauge}, one makes the following
identification: $X^m(\xi^n)=\xi^m$. It is applied to fix the gauge
freedom due to the invariance of the action (\ref{pa}) with
respect to infinitesimal diffeomorphisms (reparametrizations) of
the string worldsheet. Instead, we will use a more general gauge
than the static one. It exploits the symmetry of the background,
which exists for every physically relevant external fields.
Namely, our ansatz for the string coordinates $X^M=(X^\mu, X^a)$
is given by (\ref{tLA}), and $x^\mu$ are the target space-time
coordinates, on which the background fields do not depend:
\ba\label{ob}\p_\mu g_{MN} =0,\h \p_\mu b_{MN} =0.\ea If we
restrict ourselves to $\mu=m$ and $\Lambda^m_n=\delta^m_n$, we
come back to static gauge \footnote{We note however that this
gauge has been never used when semi-classical string quantization
was considered.}.

Taking into account the ansatz (\ref{tLA}), the Lagrangian
density, the induced metric, the constraints (\ref{0}) and
(\ref{1}) respectively, and the Euler-Lagrange equations for $X^M$
(\ref{sem}), can be written as (the over-dot is used for
$d/d\tau$)\ba \label{LRa} \mathcal{L}^{A}(\tau) =
-\frac{T}{2}\sqrt{- \gamma}\left[ \gamma^{00}g_{ab}\dot{Y}^a
\dot{Y}^b+ 2\left( \gamma^ {0n}g_{a \nu}\Lambda_n^\nu -
\frac{1}{\sqrt{- \gamma}}\Lambda_1^\nu b_{a \nu} \right)\dot{Y}^a
\right.+
\\ \nl \left. + \gamma^{mn}\Lambda_m^\mu \Lambda_n^\nu g_{\mu \nu}-
\frac{2}{\sqrt{-\gamma}}\Lambda_0^\mu \Lambda_1^\nu b_{\mu \nu}
\right];\ea \ba\label{imtLA} &&G_{00}=g_{ab}\dot{Y}^a\dot{Y}^b +
2\Lambda^\nu_0g_{\nu a}\dot{Y}^a +
\Lambda^\mu_0\Lambda^\nu_0g_{\mu\nu},\\ \nl
&&G_{01}=\Lambda^\nu_1\left(g_{\nu a}\dot{Y}^a +
\Lambda^\mu_0g_{\mu\nu}\right),\h
G_{11}=\Lambda^\mu_1\Lambda^\nu_1g_{\mu\nu};\ea \ba\label{a0}
&&\gamma^{00}g_{ab}\dot{Y}^a\dot{Y}^b +
2\gamma^{00}\Lambda^\nu_0g_{\nu a}\dot{Y}^a +
\left(\gamma^{00}\Lambda^\mu_0\Lambda^\nu_0 -
\gamma^{11}\Lambda^\mu_1\Lambda^\nu_1\right)g_{\mu\nu}=0,\\
\label{a1} &&\Lambda^\nu_1\left(\gamma^{00}g_{\nu a}\dot{Y}^a +
\gamma^{0n} \Lambda^\mu_n g_{\mu\nu}\right)=0;\ea \ba\label{aem}
&&\gamma^{00}\left(g_{Lb}\ddot{Y}^b +
\Gamma_{L,bc}\dot{Y}^b\dot{Y}^c \right) +
2\gamma^{0n}\Lambda^\mu_n\Gamma_{L,\mu b}\dot{Y}^b +
\gamma^{mn}\Lambda^{\mu}_{m}\Lambda^{\nu}_{n}\Gamma_{L,\mu\nu} \\
\nl &&=-\frac{1}{\sqrt{- \gamma}}\Lambda_1^\nu \left( H_{L \mu
\nu}\Lambda_0^\mu + H_{L a \nu} \dot{Y}^a \right).\ea $
\mathcal{L}^{A}(\tau)$ in (\ref{LRa}) is like a Lagrangian for a
point particle, interacting with the external fields $g_{MN}$,
$b_{a\nu}$ and $b_{\mu\nu}$.

Let us write down the conserved quantities. By definition, the
generalized momenta are \ba \nl P_L\equiv\frac{\p \mathcal{L}}{\p
(\p_0 X^L)} = -T \left(\sqrt{-\gamma}\gamma ^{0n} g_{LN}\p_n X^N -
b_{LN} \p_1 X^N \right). \ea For our ansatz, they take the form:
\ba\nl P_L = -T \left[\sqrt{- \gamma}\left( \gamma^{00} g_{La}
\dot{Y}^a+ \gamma ^{0n}g_{L \nu}\Lambda_n^\nu \right)- b_{L
\nu}\Lambda ^\nu_1 \right].\ea The Lagrangian (\ref{LRa}) does not
depend on the coordinates $X^\mu$. Therefore, the conjugated
momenta $P_\mu$ are conserved \ba\nl  P_\mu = -T \left[\sqrt{-
\gamma}\left( \gamma^{00} g_{\mu a} \dot{Y}^a+ \gamma
^{0n}\Lambda_n^\nu g_{\mu \nu}\right)-\Lambda ^\nu_1
b_{\mu\nu}\right]= constants.\ea

The same result can be obtained by solving the equations of motion
(\ref{aem}) for $L=\lambda$. In accordance with (\ref{ob}), the
computation of $\Gamma_{\lambda,MN}$ and $H_{\lambda,MN}$ gives
\ba\nl &&\Gamma_{\lambda,ab}=\frac{1}{2}\left(\p_ag_{b\lambda}
+\p_bg_{a\lambda}\right), \h \Gamma_{\lambda,\mu
a}=\frac{1}{2}\p_ag_{\mu\lambda},\h \Gamma_{\lambda,\mu\nu}=0,\\
\nl &&H_{\lambda ab}=\p_a b_{b\lambda} + \p_b b_{\lambda a},\h
H_{\lambda\mu a}=\p_a b_{\lambda\mu},\h H_{\lambda\mu\nu}=0.\ea
Inserting these expressions in the part of the differential
equations (\ref{aem}) corresponding to $L=\lambda$, and using the
equalities $\dot{g}_{MN}=\dot{Y}^a\p_ag_{MN}$,
$\dot{b}_{MN}=\dot{Y}^a\p_a b_{MN}$, one receives the first
integrals \ba\nl \gamma^{00}g_{\lambda a}\dot{Y}^a +
\gamma^{0n}\Lambda^{\mu}_{n}g_{\lambda\mu} - \frac{1}{\sqrt{-
\gamma}}\Lambda_1 ^\nu b_{\lambda\nu}= constants.\ea  It is easy
to check that they are connected with the conserved momenta
$P_\mu$ as \ba\label{cm} \gamma^{00}g_{\mu a}\dot{Y}^a +
\gamma^{0n}\Lambda^{\nu}_{n}g_{\mu\nu} - \frac{1}{\sqrt{-
\gamma}}\Lambda_1 ^\nu b_{\mu\nu} =
-\frac{P_\mu}{T\sqrt{-\gamma}}.\ea

From (\ref{a1}) and (\ref{cm}), one obtains the following
compatibility condition \ba\label{cc} \Lambda^{\nu}_{1}P_\nu =
0.\ea This equality may be interpreted as a solution of the
constraint (\ref{a1}), which restricts the number of the
independent parameters in the theory.

With the help of (\ref{cm}), the other constraint, (\ref{a0}), can
be rewritten in the form \ba\label{ec} g_{ab}\dot{Y}^a\dot{Y}^b =
\mathcal{U},\ea where $\mathcal{U}$ is given by \ba\label{sp}
\mathcal{U}=\frac{1}{\gamma^{00}} \left[
\gamma^{mn}\Lambda^{\mu}_{m}\Lambda^{\nu}_{n}g_{\mu\nu} +
\frac{2\Lambda^{\mu}_{0}}{T\sqrt{-\gamma}} \left(P_\mu- T
\Lambda_1^\nu b_{\mu \nu}\right) \right].\ea

Now, let us turn to the equations of motion (\ref{aem}),
corresponding to $L=a$. By using the explicit expressions \ba\nl
&&\Gamma_{a,\mu
b}=-\frac{1}{2}\left(\p_ag_{b\mu}-\p_bg_{a\mu}\right)
=-\p_{[a}g_{b]\mu},\h
\Gamma_{a,\mu\nu}=-\frac{1}{2}\p_ag_{\mu\nu},\\ \nl &&
H_{a\mu\nu}= \p_a b_{\mu\nu}; \h H_{ab\nu}=\p_ab_{b\nu} - \p_b
b_{a \nu}= 2\p_{[a}b_{b]\nu},\ea one obtains \ba\label{fem}
g_{ab}\ddot{Y}^b + \Gamma_{a,bc}\dot{Y}^b\dot{Y}^c =
\frac{1}{2}\p_a \mathcal{U} +
2\p_{[a}\mathcal{A}_{b]}\dot{Y}^b.\ea In (\ref{fem}), an effective
potential $\mathcal{U}$ and an effective gauge field
$\mathcal{A}_a$ appeared. $\mathcal{U}$ is given in (\ref{sp}),
and \ba\label{gf} \mathcal{A}_a= \frac{1}{\gamma
^{00}}\left(\gamma^{0 m}\Lambda_m^\mu g_{a\mu}-
\frac{\Lambda_1^\mu b_{a \mu}}{\sqrt{-\gamma}} \right).\ea

The reduced equations of motion (\ref{fem}) are as for a point
particle moving in the gravitational field $g_{ab}$, in the
potential $\mathcal{U}$ and interacting with the 1-form gauge
field $\mathcal{A}_a$ through its field strength
$\mathcal{F}_{ab}=2\p_{[a}\mathcal{A}_{b]}$.

Now our task is to find {\it exact} solutions of the {\it
nonlinear} differential equations (\ref{ec}) and (\ref{fem}). It
turns out that for background fields depending on only one
coordinate $x^a$, we can always integrate these equations, and the
solution is \footnote{In this case, the constraint (\ref{ec}) is
first integral for the equation of motion (\ref{fem}).}
\ba\label{ocs}\tau\left(X^a\right)=\tau_0 \pm \int_{X_0^a}^{X^a}d
x \left(\frac{\mathcal{U}}{g_{aa}}\right)^{-1/2}.\ea Otherwise,
supposing the metric $g_{ab}$ is a diagonal one, (\ref{fem}) and
(\ref{ec}) reduce to \ba\label{dem}
&&\frac{d}{d\tau}(g_{aa}\dot{Y}^a) -\frac{1}{2}\left[\p_a g_{aa}(
\dot{Y}^a)^2 +\p_a\mathcal{U}\right]-\frac{1}{2}\sum_{b\ne
a}\left[\p_a g_{bb}(\dot{Y}^b)^2 +
4\p_{[a}\mathcal{A}_{b]}\dot{Y}^b\right] = 0,\\ \label{dec}
&&g_{aa}(\dot{Y}^a)^2+\sum_{b\ne a}
g_{bb}(\dot{Y}^b)^2=\mathcal{U}.\ea With the help of the
constraint (\ref{dec}), we can rewrite the equations of motion
(\ref{dem}) in the form \ba\label{st}
\frac{d}{d\tau}(g_{aa}\dot{Y}^a)^2 - \dot{Y}^a\p_a\left(g_{aa}
\mathcal{U}\right)+ \dot{Y}^a\sum_{b\ne a}
\left[\p_a\left(\frac{g_{aa}}{g_{bb}}\right) (g_{bb}\dot{Y}^b)^2 -
4g_{aa}\p_{[a}\mathcal{A}_{b]} \dot{Y}^b\right] = 0.\ea

To find solutions of the above equations without choosing
particular background, we can fix all coordinates $Y^a$ except
one. Then the $exact$ string solution of the equations of motion
and constraints is given again by the same expression (\ref{ocs})
for $\tau\left(X^a\right)$.

To find solutions depending on more than one coordinate, we have
to impose further conditions on the background fields. Let us
first consider the simpler case, when the last two terms in
(\ref{st}) are not present. This may happen, when \ba\label{sc}
\p_a\left(\frac{g_{aa}}{g_{bb}}\right) = 0,\h
\mathcal{A}_{a}=0.\ea Then, the first integrals of (\ref{st}) are
\ba\label{fias} \left(g_{aa}\dot{Y}^a\right)^2 =D_a(Y^{b\ne
a})+g_{aa}\mathcal{U},\ea where $D_a$ are arbitrary functions of
their arguments. These solutions must be compatible with the
constraint (\ref{dec}), which leads to the condition
\ba\nl\sum_a\frac{D_a}{g_{aa}}=(1-n_a)\mathcal{U},\ea where $n_a$
is the number of the coordinates $Y^a$. From here, one can express
one of the functions $D_a$ through the others. To this end, we
split the index $a$ in such a way that $Y^r$ is one of the
coordinates $Y^a$, and $Y^{\alpha}$ are the others. Then \ba\nl
D_r=-g_{rr}\left(n_{\alpha}\mathcal{U}
+\sum_{\alpha}\frac{D_{\alpha}}{g_{\alpha\alpha}}\right),\ea and
by using this, one rewrites the first integrals (\ref{fias}) as
\ba\label{fiasf} \left(g_{rr}\dot{Y}^r\right)^2 = g_{rr}
\left[(1-n_{\alpha})\mathcal{U} - \sum_{\alpha}
\frac{D_{\alpha}}{g_{\alpha\alpha}}\right]\ge 0,\h
\left(g_{\alpha\alpha}\dot{Y}^\alpha\right)^2 =D_\alpha(Y^{a\ne
\alpha})+g_{\alpha\alpha}\mathcal{U}\ge 0, \ea where $n_{\alpha}$
is the number of the coordinates $Y^\alpha$. Thus, the constraint
(\ref{dec}) is satisfied identically.

Now we turn to the general case, when all terms in the equations
of motion (\ref{st}) are present. The aim is to find conditions,
which will allow us to reduce the order of the equations of motion
by one. An example of such {\it sufficient} conditions, is given
below : \ba\nl &&\mathcal{A}_a\equiv\left(
\mathcal{A}_r,\mathcal{A}_{\alpha} \right)= \left(
\mathcal{A}_r,\p_{\alpha}f \right),\h
\p_{\alpha}\left(\frac{g_{\alpha\alpha}}{g_{aa}}\right)=0,\\ \nl
&&\p_{\alpha}\left(g_{rr}\dot{Y}^r\right)^2 = 0,\h
\p_{r}\left(g_{\alpha\alpha}\dot{Y}^{\alpha}\right)^2 = 0.\ea By
using the restrictions given above, one obtains the following
first integrals of the equations (\ref{st}), compatible with the
constraint (\ref{dec}) \ba\label{fir}
\left(g_{rr}\dot{Y}^{r}\right)^2 &=&
g_{rr}\left[\left(1-n_{\alpha}\right) \mathcal{U} -
\sum_{\alpha}\frac{D_{\alpha}}{g_{\alpha\alpha}} -
2n_{\alpha}\left(\mathcal{A}_{r}-\p_r f\right)\dot{Y}^{r}\right]=
E_r\left(Y^r\right)\ge 0,\\ \label{fia}
\left(g_{\alpha\alpha}\dot{Y}^{\alpha}\right)^2 &=& D_{\alpha}
\left(Y^{a\ne\alpha}\right) +
g_{\alpha\alpha}\left[\mathcal{U}+2\left( \mathcal{A}_{r}-\p_r
f\right)\dot{Y}^r\right]= E_{\alpha}\left(Y^{\beta}\right)\ge 0,
\ea where $D_{\alpha}$, $E_{\alpha}$ and $E_r$ are arbitrary
functions of their arguments.

Further progress is possible, when working with particular
background configurations, allowing for separation of the
variables in (\ref{fiasf}), or in (\ref{fir}) and (\ref{fia}).

\subsection{The tensionless limit}
Our results obtained so far are not applicable to tensionless
(null) strings, because the action (\ref{pa}) is proportional to
the string tension $T$. The parameterization of $\gamma^{mn}$,
which is appropriate for considering the zero tension limit $T\to
0$, is the following \cite{ILST93, HLU94}: \ba\label{tl}
\gamma^{00}=-1,\h \gamma^{01}=\lambda^1,\h
\gamma^{11}=(2\lambda^0T)^2 - (\lambda^1)^2, \h \det(\gamma^{mn})=
-(2\lambda^0T)^2.\ea Here $\lambda^n$ are the Lagrange
multipliers, whose equations of motion generate the {\it
independent} constraints. In these notations, the constraints
(\ref{a0}) and (\ref{a1}), the equations of motion (\ref{aem}),
and the conserved momenta (\ref{cm}) take the form \ba\nl
&&g_{ab}\dot{Y}^a\dot{Y}^b + 2\Lambda^\nu_0g_{\nu a}\dot{Y}^a +
\left\{\Lambda^\mu_0\Lambda^\nu_0
+\left[(2\lambda^0T)^2-(\lambda^1)^2\right]
\Lambda^\mu_1\Lambda^\nu_1\right\} g_{\mu\nu}=0,\\ \nl
&&\Lambda^\nu_1\left[g_{\nu a}\dot{Y}^a +
\left(\Lambda^\mu_0-\lambda^1\Lambda^\mu_1\right)
g_{\mu\nu}\right]=0;\ea \ba\nl &&g_{Lb}\ddot{Y}^b +
\Gamma_{L,bc}\dot{Y}^b\dot{Y}^c +
2\left(\Lambda^\mu_0-\lambda^1\Lambda^\mu_1\right)\Gamma_{L,\mu
b}\dot{Y}^b\\ \nl
&&+\left[\left(\Lambda^\mu_0-\lambda^1\Lambda^\mu_1\right)
\left(\Lambda^\nu_0-\lambda^1\Lambda^\nu_1\right)-(2\lambda^0T)^2
\Lambda^\mu_1\Lambda^\nu_1\right]\Gamma_{L,\mu\nu}= 2 \lambda^0 T
\Lambda_1^\nu \left(H_{L \nu b}\dot{Y}^b + \Lambda_0^\mu H_{L\mu
\nu}\right);\ea \ba\nl g_{\mu a}\dot{Y}^a +
\left(\Lambda^\nu_0-\lambda^1\Lambda^\nu_1\right)g_{\mu\nu} +
2\lambda^0T\Lambda^\nu_1b_{\mu\nu}= 2\lambda^0P_\mu.\ea The
reduced equations of motion and constraint (\ref{fem}) and
(\ref{ec}) have the same form, but now, the effective potential
(\ref{sp}) and the effective gauge field (\ref{gf}) are given by
\ba\nl &&\mathcal{U}^{\lambda}=
\left[\left(\Lambda^\mu_0-\lambda^1\Lambda^\mu_1\right)
\left(\Lambda^\nu_0-\lambda^1\Lambda^\nu_1\right)-(2\lambda^0T)^2
\Lambda^\mu_1\Lambda^\nu_1\right]g_{\mu\nu}
-4\lambda^0\Lambda_0^\mu\left( P_\mu - T\Lambda_1^\nu
b_{\mu\nu}\right),\\ \nl &&\mathcal{A}^{\lambda}_a
=\left(\Lambda^\mu_0-\lambda^1\Lambda^\mu_1\right) g_{a\mu} + 2
\lambda^0 T\Lambda_1^\mu b_{a \mu}.\ea

If one sets $\lambda^1=0$ and $2\lambda^0T=1$, the results in {\it
conformal gauge}  are obtained, as it should be. If one puts $T=0$
in the above formulas, they will describe {\it tensionless}
strings.

\section{Exact solutions for more general string embedding}
Now, we are going to use the ansatz (\ref{tGA}) for the string
coordinates, which corresponds to more general string embedding.
Here, compared with (\ref{tLA}), $X^\mu$ are allowed to vary
non-linearly with the proper time $\tau$. In addition, we assume
that the conditions (\ref{ob}) on the background fields still
hold.

By using the ansatz (\ref{tGA}), one obtains that the Lagrangian
density, the induced metric, the constraints (\ref{0}) and
(\ref{1}) respectively, and the Euler-Lagrange equations for $X^M$
(\ref{sem}) are given by \ba\nl \mathcal{L}^{GA}(\tau) =
-\frac{T}{2}\sqrt{-\gamma}\left[\gamma^{00}g_{MN} \dot{Y}^M
\dot{Y}^N + 2 \left(\gamma^{0n}\Lambda_n^\nu g_{M \nu}-
\frac{\Lambda_1^\nu b_{M \nu}}{\sqrt{- \gamma}} \right)\dot{Y}^M +
\right.\\ \nl +\left. \gamma^{mn}\Lambda_m^\mu \Lambda_n^\nu
g_{\mu \nu}-\frac{2\Lambda _0^\mu\Lambda_1^\nu
b_{\mu\nu}}{\sqrt{-\gamma}} \right];\ea \ba \label{imga}
&&G_{00}=g_{MN}\dot{Y}^M\dot{Y}^N + 2\Lambda^\nu_0g_{\nu
N}\dot{Y}^N + \Lambda^\mu_0\Lambda^\nu_0g_{\mu\nu},\\ \nl
&&G_{01}=\Lambda^\nu_1\left(g_{\nu N}\dot{Y}^N +
\Lambda^\mu_0g_{\mu\nu}\right),\h
G_{11}=\Lambda^\mu_1\Lambda^\nu_1g_{\mu\nu};\ea \ba\label{mga0}
&&\gamma^{00}g_{MN}\dot{Y}^M\dot{Y}^N +
2\gamma^{00}\Lambda^\nu_0g_{\nu N}\dot{Y}^N +
\left(\gamma^{00}\Lambda^\mu_0\Lambda^\nu_0 -
\gamma^{11}\Lambda^\mu_1\Lambda^\nu_1\right)g_{\mu\nu}=0,\\
\label{mga1} &&\Lambda^\nu_1\left(\gamma^{00}g_{\nu N}\dot{Y}^N +
\gamma^{0n} \Lambda^\mu_n g_{\mu\nu}\right)=0;\ea \ba\label{mgaem}
\gamma^{00}\left(g_{LN}\ddot{Y}^N +
\Gamma_{L,MN}\dot{Y}^M\dot{Y}^N \right) +
2\gamma^{0n}\Lambda^\mu_n\Gamma_{L,\mu N}\dot{Y}^N +
\gamma^{mn}\Lambda^{\mu}_{m}\Lambda^{\nu}_{n}\Gamma_{L,\mu\nu}= \\
\nl = -\frac{1}{\sqrt{- \gamma}}\Lambda_1 ^\nu\left(H_{LM \nu}
\dot{Y}^M + \Lambda_0^\mu H_{L \mu\nu}\right) .\ea The conserved
momenta $P_\mu$ can be found as before, and now they are
\ba\label{mgcm} \gamma^{00}g_{\mu N}\dot{Y}^N +
\gamma^{0n}\Lambda^{\nu}_{n}g_{\mu\nu}- \frac{\Lambda_1^\nu
b_{\mu\nu}}{\sqrt{- \gamma}} = -\frac{P_\mu}{T\sqrt{-\gamma}}=
constants.\ea The compatibility condition following from the
constraint (\ref{mga1}) and from (\ref{mgcm}) coincides with the
previous one (\ref{cc}). With the help of (\ref{mgcm}), the
equations of motion (\ref{mgaem}) corresponding to $L=a$ and the
other constraint (\ref{mga0}), can be rewritten in the form
\ba\label{mgmem} &&g_{aN}\ddot{Y}^N +
\Gamma_{a,MN}\dot{Y}^M\dot{Y}^N = \frac{1}{2}\p_a \mathcal{U} +
2\p_{[a}\mathcal{A}_{N]}\dot{Y}^N,\\ \label{mgec}
&&g_{MN}\dot{Y}^M\dot{Y}^N = \mathcal{U},\ea where $\mathcal{U}$
is given by (\ref{sp}) and \ba\label{mggm} \mathcal{A}_N=
\frac{1}{\gamma ^{00}}\left(\gamma^{0 m}\Lambda_m^\mu g_{N\mu}-
\frac{\Lambda_1^\mu b_{N \mu}}{\sqrt{-\gamma}} \right) \ea
coincides with (\ref{gf}) for $N=a$.

Till now, all is much like before, and putting $\dot{Y}^\mu=0$ in
the equalities based on the new ansatz (\ref{tGA}), we will
retrieve those, which follow from the previous one ((\ref{tLA})).

Now we are going to eliminate the variables $\dot{Y}^\mu$ from
(\ref{mgmem}) and (\ref{mgec}). To this end, we express
$\dot{Y}^\mu$ through $\dot{Y}^a$ from the conservation laws
(\ref{mgcm}): \ba\label{muc} \dot{Y}^\mu =
-\frac{\gamma^{0n}}{\gamma^{00}}\Lambda^\mu_n -
\left(g^{-1}\right)^{\mu\nu}\left[g_{\nu a}\dot{Y}^a +
\frac{1}{\gamma^{00}T\sqrt{-\gamma}}\left( P_\nu - T\Lambda_1^\rho
b_{\nu \rho} \right)\right].\ea

After using (\ref{muc}) and (\ref{cc}), the equations of motion
(\ref{mgmem}) and the constraint (\ref{mgec}) acquire the form
\ba\label{mgemf} &&h_{ab}\ddot{Y}^b +
\Gamma^{\bf{h}}_{a,bc}\dot{Y}^b\dot{Y}^c = \frac{1}{2}\p_a
\mathcal{U}^{\bf{h}} +
2\p_{[a}\mathcal{A}^{\bf{h}}_{b]}\dot{Y}^b,\\ \label{mgecf}
&&h_{ab}\dot{Y}^a\dot{Y}^b = \mathcal{U}^{\bf{h}},\ea where a new,
effective metric appeared \ba\nl h_{ab} = g_{ab} -
g_{a\mu}(g^{-1})^{\mu\nu}g_{\nu b}.\ea $\Gamma^{\bf{h}}_{a,bc}$ is
the symmetric connection corresponding to this metric \ba\nl
\Gamma^{\bf{h}}_{a,bc}=\frac{1}{2}\left(\p_bh_{ca}
+\p_ch_{ba}-\p_ah_{bc}\right).\ea The new effective scalar and
gauge potentials, expressed through the background fields, are as
follows \ba\nl
&&\mathcal{U}^{\bf{h}}=\frac{1}{\gamma\left(\gamma^{00}\right)^2}
\left[ \Lambda^{\mu}_{1}\Lambda^{\nu}_{1}g_{\mu\nu}+ \frac{1}{T^2}
\left( P_\mu - T\Lambda_1^\rho b_{\mu \rho}\right)
(g^{-1})^{\mu\nu} \left( P_\nu - T\Lambda_1^\lambda
b_{\nu\lambda}\right)\right],
\\ \nl &&\mathcal{A}^{\bf{h}}_{a}= -
\frac{1}{\gamma^{00}T\sqrt{-\gamma}} \left[g_{a\mu}
(g^{-1})^{\mu\nu} \left(P_\nu - T\Lambda _1 ^ \rho b_{\nu\rho}
\right)+ T \Lambda_1 ^\rho b_{a \rho} \right] .\ea We point out
the qualitatively different behaviour of the potentials
$\mathcal{U}^{\bf{h}}$ and $\mathcal{A}_a^{\bf{h}}$, compared to
$\mathcal{U}$ and $\mathcal{A}_a$, due to the appearance of the
inverse metric $(g^{-1})^{\mu\nu}$.

Since the equations (\ref{fem}), (\ref{ec}) and (\ref{mgemf}),
(\ref{mgecf}) have the same form, for obtaining exact string
solutions, we  can proceed as before and use the previously
derived formulas after the replacements
$(g,\Gamma,\mathcal{U},\mathcal{A})$ $\to$ $(h,\Gamma^{\bf{h}},
\mathcal{U}^{\bf{h}},\mathcal{A}^{\bf{h}})$. In particular, the
solution depending on one of the coordinates $X^a$ will be
\ba\label{mgocs}\tau\left(X^a\right)=\tau_0 \pm
\int_{X_0^a}^{X^a}d x
\left(\frac{\mathcal{U}^{\bf{h}}}{h_{aa}}\right)^{-1/2}.\ea In
this case by integrating (\ref{muc}), and replacing the solution
for $Y^\mu$ in the ansatz (\ref{tGA}), one obtains the solution
for the string coordinates $X^\mu$: \ba \label{X} &&X^\mu(X^a,
\sigma) = X_0^\mu + \Lambda_1 ^\mu \left[ \sigma -
\frac{\gamma^{01}}{\gamma^{00}}\tau\left(X^a\right) \right] -\\
\nl && - \int_ {X_0^a}^ {X^a} (g^{-1})^{\mu\nu} \left[g_{\nu a}\pm
\frac{\left(P_\nu - T\Lambda_1^\rho b_{\nu\rho}\right)}
{\gamma^{00}T \sqrt{- \gamma}} \left(\frac{\mathcal
{U}^{\rm{h}}}{h_{aa}} \right)^{-1/2} \right] d x .\ea

To be able to take the tensionless limit $T\to 0$ in the above
formulas, we have to use the $\lambda$-parameterization (\ref{tl})
of $\gamma^{mn}$. The quantities that appear in the reduced
equations of motion and constraint (\ref{mgemf}) and
(\ref{mgecf}), which depend on this parameterization, are
$\mathcal{U}^{\bf{h}}$ and $\mathcal{A}^{\bf{h}}_{a}$. Now, they
are given by \ba\nl && \mathcal{U}^{h,\lambda}= - (2\lambda^0)^2
\left[T^2 \Lambda_1^\mu \Lambda_1 ^\nu g_{\mu\nu} + \left(P_\mu -
T\Lambda_1^\rho b_{\mu\rho} \right) (g^{-1})^{\mu\nu}
\left(P_\nu-T\Lambda_1^\lambda b_{\nu\lambda}\right)\right], \\
\nl && \mathcal{A}_{a}^{h,\lambda} = 2\lambda^0
\left[g_{a\mu}(g^{-1})^{\mu\lambda} (P_\lambda - T\Lambda_1^\rho
b_{\lambda\rho}) + T\Lambda_1^\rho b_{a\rho} \right]. \ea If one
sets $\lambda^1=0$ and $2\lambda^0T=1$, the {\it conformal gauge}
results are obtained. If one puts $T=0$ in the above equalities,
they will correspond to {\it tensionless} strings.

\section{Some applications}
In the previous two sections, we described a general approach for
solving the string equations of motion and constraints in the
background fields $g_{MN}(x)$ and $b_{MN}(x)$, with the help of
the conserved momenta $P_\mu$, based on the ansatzes (\ref{tLA})
and (\ref{tGA}). In this section, as an illustration of the
previously obtained general results, we will establish the
correspondence with the particular cases considered in \cite{11}
in the framework of the linear ansatz (\ref{tLA}), and in \cite{4}
- in the framework of the nonlinear ansatz (\ref{tGA}).

In \cite{11}, the string theory background is $AdS_5\times S^5$,
with field theory dual $\mathcal{N}=4$ $SU(N)$ $SYM$ in four
dimensional flat space-time. Two cases was considered: pulsating
strings in $AdS_5$ and on $S^5$. The metric of the  $AdS_5$ is
taken to be \ba\nl &&ds^2_{AdS_5}= R^2\left(-\cosh^2\rho dt^2 +
d\rho^2 + \sinh^2\rho d\Omega^2_3\right),\\ \nl &&d\Omega^2_3=
\cos^2\theta d\psi^2 + d\theta^2 + \sin^2\theta d\phi^2,\h
R^4=\lambda\alpha'^2.\ea The metric on the $S^5$ is given by
\ba\nl ds^2_{S^5}=R^2 \left(d\theta^2_1 + \sin^2\theta_1 d\psi^2_1
+ \cos^2\theta_1 d\Omega^{'2}_{3}\right).\ea

For the pulsating circular string in $AdS_5$, the following ansatz
has been used \ba\label{pa1} t=\tau,\h \rho=\rho(\tau),\h
\phi=m\sigma,\h \theta=\pi/2.\ea The relevant metric seen by the
string is \ba\label{rm1} ds^2=R^2\left(-\cosh^2\rho dt^2 + d\rho^2
+ \sinh^2\rho d\phi^2\right).\ea Therefore, $b_{MN}=0$ and this
metric does not depend on $x^0=t$ and $x^2=\phi$, i.e.
$X^\mu=X^{0,2}$ and $X^a=X^1$ in our notations. Comparing the
ansatzes (\ref{tLA}) and (\ref{pa1}), one can see that the latter
is particular case of the former, corresponding to \ba\nl
\Lambda_0^0=1,\h \Lambda^0_1=\Lambda_0^2=0,\h \Lambda_1^2=m.\ea
The induced metric (\ref{imtLA}) is \ba\nl G_{00}=R^2\left(
\dot{\rho}^2-\cosh^2\rho\right),\h G_{01}=0,\h
G_{11}=m^2R^2\sinh^2\rho.\ea Taking this into account, one
reproduces the Nambu-Goto action, used in \cite{11}, for the case
under consideration: \ba\nl S=-m\sqrt{\lambda}\int dt
\sinh\rho\sqrt{\cosh^2\rho-\dot{\rho}^2}.\ea

The conserved energy, obtained from (\ref{cm}), is \ba\nl E=-2\pi
P_0= -2\pi TR^2\sqrt{-\gamma}\gamma^{00}\cosh^2\rho =
-\sqrt{-\lambda\gamma}\gamma^{00}\cosh^2\rho.\ea

The background metric (\ref{rm1}) depends on only one coordinate,
so our string solution is given by (\ref{ocs}): \ba\nl
\tau(\rho)=\tau_0 \pm
\sqrt{-\gamma^{00}}\int_{\rho_0}^{\rho}\frac{d\rho}
{\sqrt{\frac{2E}{\sqrt{-\lambda\gamma}} + \gamma^{00}\cosh^2\rho -
\gamma^{11}m^2\sinh^2\rho}}.\ea In {\it conformal gauge}, this
solution takes the form \ba\nl \tau(\rho)=\tau_0 \pm
\int_{\rho_0}^{\rho}\frac{d\rho}{\sqrt{\frac{2E}{\sqrt{\lambda}}
-\left(\cosh^2\rho + m^2\sinh^2\rho\right)}}.\ea In the {\it
tensionless} limit, one obtains:\ba\nl \tau(\rho)_{T=0}=\tau_0 \pm
\int_{\rho_0}^{\rho}\frac{d\rho} {\sqrt{- \cosh^2\rho +\left(
\lambda^1 \right)^2m^2\sinh^2\rho}}.\ea

The second case considered in \cite{11} is based on the ansatz:
\ba\label{pa2} t=\tau,\h \rho=\rho(\tau),\h
\theta_1=\theta_1(\tau),\h \psi_1=m\sigma.\ea The relevant metric
is \ba\label{rm2} ds^2=R^2\left(-\cosh^2\rho dt^2 + d\rho^2 +
d\theta_1^2 + \sin^2\theta_1 d\psi_1^2\right)\ea and it does not
depend on $x^0=t$ and $x^3=\psi_1$. Hence in our notations
$X^\mu=X^{0,3}$, $X^a=X^{1,2}$ and \ba\nl \Lambda_0^0=1,\h
\Lambda^0_1=\Lambda_0^3=0,\h \Lambda_1^3=m.\ea The induced metric
(\ref{imtLA}) is \ba\nl G_{00}=R^2\left( \dot{\rho}^2 +
\dot{\theta_1}^2 -\cosh^2\rho\right),\h G_{01}=0,\h
G_{11}=m^2R^2\sin^2\theta_1.\ea Taking this into account, one
reproduces the Nambu-Goto action, used in \cite{11}, for the case
at hand: \ba\nl S=-m\sqrt{\lambda}\int dt
\sin\theta_1\sqrt{\cosh^2\rho-\dot{\rho}^2-\dot{\theta_1}^2}.\ea

In accordance with our general considerations in Sec.2, we can
give three types of string solutions: when $\theta_1$ is fixed,
when $\rho$ is fixed, and without fixing any of the coordinates
$\theta_1$ and $\rho$, on which the background depends.

If we fix $\theta_1=\theta_1^0=constant$, the solution (\ref{ocs})
gives \ba\nl \tau(\rho)=\tau_0 \pm
\sqrt{-\gamma^{00}}\int_{\rho_0}^{\rho}\frac{d\rho}
{\sqrt{\frac{2E}{\sqrt{-\lambda\gamma}} + \gamma^{00}\cosh^2\rho -
\gamma^{11}m^2\sin^2\theta_1^0}}.\ea

If we fix $\rho=\rho_0=constant$, the solution (\ref{ocs}) is
\ba\nl \tau(\theta_1)=\tau_0 \pm
\sqrt{-\gamma^{00}}\int_{\theta_1^0}^{\theta_1}\frac{d\theta_1}
{\sqrt{\frac{2E}{\sqrt{-\lambda\gamma}} + \gamma^{00}\cosh^2\rho_0
- \gamma^{11}m^2\sin^2\theta_1}}.\ea

When none of the coordinates $\rho$ and $\theta_1$ is kept fixed,
it turns out that the conditions (\ref{sc}) on the background are
fulfilled, and therefore, the solution for the two first integrals
is given by (\ref{fiasf}): \ba\label{fia2}
\left(R^2\dot{\rho}\right)^2= -D_2(\rho)\ge 0,\h
\left(R^2\dot{\theta}_1\right)^2= D_2(\rho)+
R^2\mathcal{U}(\rho,\theta_1)\ge 0.\ea The arbitrary function
$D_2(\rho)$ can be fixed by the condition for separation of the
variables $\rho$ and $\theta_1$ in the equation for $\theta_1$.
Thus if we choose \ba\nl \frac{D_2(\rho)}{R^4}-\cosh^2\rho =
-d^2=constant,\ea then the equations (\ref{fia2}) reduce to \ba\nl
\dot{\rho}^2=d^2-\cosh^2\rho\ge 0,\h \dot{\theta}_1^2=\tilde{d}^2
+ \frac{\gamma^{11}}{\gamma^{00}}m^2\sin^2\theta_1\ge 0,\ea where
\ba\nl \tilde{d}^2= -\frac{E}{\pi TR^2 \sqrt{-\gamma} \gamma^{00}}
-d^2.\ea These equations are solved by \ba\nl \rho(\tau)=\rho_0\pm
\int_{\tau_0}^{\tau} d\tau\sqrt{d^2-\cosh^2\rho},\h
\theta_1(\tau)=\theta_1^0\pm \int_{\tau_0}^{\tau}
d\tau\sqrt{\tilde{d}^2+
\frac{\gamma^{11}}{\gamma^{00}}m^2\sin^2\theta_1}.\ea One can also
find the orbit $\rho=\rho(\theta_1)$, which is given by the
equality \ba\nl
\int_{\rho_0}^{\rho}\frac{d\rho}{\sqrt{d^2-\cosh^2\rho}}=
\pm\int_{\theta_1^0}^{\theta_1}\frac{d\theta_1}{\sqrt{\tilde{d}^2+
\frac{\gamma^{11}}{\gamma^{00}}m^2\sin^2\theta_1}}.\ea

Let us now consider a closed string, which oscillates around the
center of $AdS_5$ \cite{4}\footnote{In \cite{4}, $m=1$.}: \ba\nl
t=t(\tau),\h \rho=\rho(\tau),\h \phi=m\sigma,\h \theta=\pi/2.\ea
The metric seen by this string is the same as in (\ref{rm1}). The
difference is that the above ansatz is a particular case of our
general ansatz (\ref{tGA}), corresponding to \ba\nl
&&\Lambda_0^0=\Lambda_1^0=\Lambda_0^2=0,\h \Lambda_1^2=m;\\ \nl
&&Y^0(\tau)=t(\tau),\h Y^1(\tau)=\rho(\tau),\h Y^2(\tau)=0.\ea

The induced metric (\ref{imga}) is \ba\nl G_{00}=R^2\left(
\dot{\rho}^2-\dot{t}^2\cosh^2\rho\right),\h G_{01}=0,\h
G_{11}=m^2R^2\sinh^2\rho.\ea Taking this into account, one can
find the Nambu-Goto action, for the case under consideration:
\ba\nl S=-m\sqrt{\lambda}\int d\tau
\sinh\rho\sqrt{\dot{t}^2\cosh^2\rho-\dot{\rho}^2}.\ea

The conserved energy, obtained from (\ref{mgcm}), is \ba\nl
E=-2\pi P_0= -\frac{R^2}{\alpha'}\sqrt{-\gamma}\gamma^{00}\dot{t}
\cosh^2\rho.\ea In {\it conformal gauge}, this expression reduces
to the one given in \cite{4}.

The background metric (\ref{rm1}) depends on only one coordinate,
so our string solution is given by (\ref{mgocs}) and
(\ref{X}):\ba\nl
&&\tau(\rho)=\tau_0\pm\sqrt{-\gamma}\gamma^{00}\int_{\rho_0}^{\rho}
d\rho\left[ \frac{E^2}{\lambda\cosh^2\rho} -
\left(m\sinh\rho\right)^2\right] ^{-1/2} ,\\ \nl
&&X^0(\rho)=X^0_0\mp\int_{\rho_0}^{\rho}
\frac{d\rho}{\cosh\rho\sqrt{1 -
\left(\frac{m\sqrt{\lambda}}{E}\sinh\rho\cosh\rho\right)^2}}.\ea
In the {\it tensionless limit} one obtains: \ba\nl
\tau(\rho)_{T=0}=\tau_0\mp\frac{\pi R^2}{\lambda^0
E}\int_{\rho_0}^{\rho} d\rho\cosh\rho,\h
X^0(\rho)_{T=0}=X^0_0\mp\int_{\rho_0}^{\rho}
\frac{d\rho}{\cosh\rho}.\ea

\vspace*{.5cm} {\bf Acknowledgments} \vspace*{.2cm}

This work is supported by a Shoumen University grant under
contract {\it No.001/2003}.


\end{document}